\documentclass[a4paper,12pt]{article}

\usepackage[dvipdfmx]{graphicx}
\usepackage{times}
\usepackage{natbib}
\usepackage{here}
\usepackage{setspace}
\usepackage{multirow}
\usepackage{lmodern}
\usepackage{amsmath, amssymb}
\usepackage{bm}
\usepackage{pdflscape}
\usepackage{color}

\usepackage{booktabs}
\usepackage{afterpage}
\usepackage{textcomp}
\usepackage{float}
\usepackage{enumerate}

\makeatletter
\newcommand{\figcaption}[1]{\def\@captype{figure}\caption{#1}}
\newcommand{\tblcaption}[1]{\def\@captype{table}\caption{#1}}
\makeatother
\begin{document}
\title{Analyzing Car Thefts and Recoveries with Connections to Modeling Origin-Destination Point Patterns}
\author{Shinichiro Shirota\thanks{Corresponding author, Department of Commerce, Meiji University, Japan and Center for Advanced Intelligence Project, RIKEN, Japan. E-mail: shinichiro.shirota@gmail.com}, \quad Alan. E. Gelfand\thanks{Department of Statistical Science, Duke University, US. E-mail: alan@stat.duke.edu} \quad and \quad Jorge Mateu\thanks{Department of Mathematics, Universitat Jaume I, Spain. E-mail: mateu@mat.uji.es}}
\maketitle

\bigskip
\begin{abstract}

For a given region, we have a dataset composed of car theft locations along with a linked dataset of recovery locations which, due to partial recovery, is a relatively small subset of the set of theft locations.  For an investigator seeking to understand the behavior of car thefts and recoveries in the region, several questions are addressed.   Viewing the set of theft locations as a point pattern, can we propose useful models to explain the pattern?  What types of predictive models can be built to learn about recovery location given theft location?  Can the dependence between the point pattern of theft locations and the point pattern of recovery locations be formalized?  Can the \emph{flow} between theft sites and recovery sites be captured?

Origin-destination modeling offers a natural framework for such problems.  However, here the data is not for areal units but rather is a pair of dependent point patterns, with the recovery point pattern only partially observed.  We offer modeling approaches for investigating the questions above and apply the approaches to two datasets.  One is small from the state of Neza in Mexico with areal covariate information regarding population features and crime type. The second, a much larger one, is from Belo Horizonte in Brazil but lacks potential predictors.

\end{abstract}

\noindent%

\section{Introduction}
\label{sec:Intro}

A criminal activity which has attracted little modeling attention in the statistics literature is that of automobile thefts.
Such data will consist of a set of theft locations as points in a region, perhaps with potential predictors at areal scale.  Potential predictors include (i) demographic information such as number of individuals $15$ and older, number of individuals employed, and number of individuals with health insurance access, as well as (ii) criminal activity information such as number of burglaries and number of murders.  There will also be an associated set of recovery locations, as points, for which these potential predictors are also available.  However, recoveries typically occur for only a portion of thefts so that the set of recovery locations is only a partial set of all of the potential recovery locations.

We are motivated by two real data settings. One consists of a collection of automobile thefts, with a fraction (roughly $10\%$) of recoveries, over the state of Neza in Mexico. The data is a total of 4,016 car theft locations (after deleting
some missing locations) during 2015, over both the northern and southern parts of Neza. This dataset is small but is endowed with the foregoing areal covariate information regarding population features and crime type that can be used for explanation in our modeling strategy. See Figures 1 and 2, and Section 2 for further description. A second dataset consists of car thefts which occurred in Belo Horizonte, Brazil.  It is a much larger dataset, but lacks these potential predictors. This city is 331 km$^2$ in area and has approximately 2.4 million inhabitants. In the period from August, 1, 2000, to July, 31, 2001, the dataset consists of 5,250 \emph{pairs} of theft and recovery locations. See Figure 3 and Section 2 for a more complete description.

It is important to note limitations of the available data. The low recovery rate for the Neza dataset is disappointing.  Issues such as what happened to the $90\%$ unrecovered vehicles, how different are they from the recovered vehicles (how representative of the total thefts are the $10\%$ we have observed), and what local law enforcement might do to improve the recovery rate are evidently important and would enable us to enrich the story; unfortunately this information is not available.  The data, as provided, reflects the reality of the type of reporting for this type of crime in that region.
Furthermore, the nature of the vehicles stolen - manufacturer, condition, etc., would enable comparison of the subset of those recovered to the subset of those not recovered and might also make a promising story.  However, the only available covariates are the aggregated ones mentioned above and elaborated in Section 2.  At the level of the individual vehicle, at most we have a theft location and perhaps a subsequent recovery location.

Turning to the Belo Horizonte data, in fact there were 6,339 thefts reported during the study window with 5,250 eventually found within the city limits.  So, there is a much higher recovery rate for this data than for the Neza data (which we are unable to explain).  However, we only received the theft locations for the cars that were recovered.  It may be argued that there is potential bias in this subsample of thefts.  We cannot assess this but with nearly $85\%$ of the total thefts included, we can hope that the bias is small.  A further issue is that of a false report, e.g., the owner forgot where the vehicle was parked or the vehicle was borrowed by a friend or relative without informing the owner.  Again, intriguing inference might emerge but this information is another ``individual'' feature that is not supplied.

Acknowledging the foregoing limitations, the contribution here is to adopt the perspective of crime data analysts/investigators trying to better understand the behavior of car thefts for a specified region.
It is useful to make a distinction between spatial analysis of crime data in general and spatial analysis of vehicle theft-recovery data which is our focus.
\cite{Jacobsetal(03)} state that ``carjacking
remains an under-researched and poorly understood crime.'' These authors investigate the decision-making mechanism of active carjackers in real-life settings, and conclude that such decisions are influenced by involvement in urban street culture in a very different way from other types of crimes.
Also, \cite{Kuangetal(17)} state that car thefts behave differently in a number of aspects to many other types of crimes. Indeed, as spatio-temporal crime hotspots are often detected for a number set of crimes and crime topics \citep[see][]{AndresenMalleson(15), MallesonAndresen(15)}, clusters in the form of hotspots are more rare for car thefts.

A first issue crime data analysts might focus on would be to attempt to understand the \emph{point pattern} of car thefts.  They might seek a ``risk'' surface for theft.
In Section 3 below, we offer modeling to provide an intensity surface for the point pattern of thefts to clarify where risk is high, where it is low. Specifically, Section 3 explores how car thefts, viewed as a random point pattern, can be explained using available covariates.
Using spatial regressors is well investigated in criminology contexts and there are many methods such as risk terrain modeling \citep{CaplanKennedy(10), Kennedyetal(11), Kennedyetal(16)}, which incorporate dynamic interaction among social, behavioral and environmental factors \citep{Kennedyetal(11)} along with other regression-based approaches.
There are also various hotspot detection techniques,  based on locations of crimes events, derived from tools like kernel densities estimation and clustering \citep{Chaineyetal(08)}.

The models presented in Section 3 advance the state of the art in analyzing crime data. Regression methods have serious weaknesses, e.g., in a spatial setting they fail to incorporate an intensity function to explain locations of crimes while nonparametric hotspot methods do not include covariates and hence do not provide evidence about factors to reduce crime rates

A second issue becomes one of attempting to predict recovery location given theft location.  An effective predictive model would help local law enforcement in the process of vehicle recovery.  In Section 4 below, we offer modeling to provide such prediction.

A third issue, taken up in Section 5, is spatial interaction/origin-destination modeling.  In the literature, such modeling is customarily developed at areal scale.  That is, the study region is partitioned into municipal units, e.g., postcodes, census units, business districts, labor markets.  The observations are typically counts associated with a pair of areal units, an origin unit and a destination unit.  An example is the number of individuals living in origin unit $i$ and working in destination unit $j$. In addition, we would have potential regressors associated with each of the areal units as well as a suitable \emph{distance} between the units, e.g., a road distance or a commuting time. The origin-destination modeling obtains $\{p_{ij}\}$, the matrix of origin-destination probabilities, e.g., the probability of living in unit $i$ and working in unit $j$ \citep{Chakrabortyetal(13)}, or the probability of a mail originating from unit $i$ and sent to unit $j$ \citep{Banerjeeetal(00)}.  Interest lies in \emph{flows}, the number of people who live in unit $i$ and work in unit $j$, e.g., $n_{i}p_{ij}$ where $n_{i}$ is the number of people living in unit $i$.


Within this origin-destination framework, our car theft setting differs from the above in two ways.  First, the data is available at point level and can be viewed as a pair of dependent point patterns.  Second, the recovery point pattern is typically only partially observed.  When a complete pair is observed, we have a geo-coded origin location and a geo-coded destination location; when recovery is missing, we have only a geo-coded origin.  Regardless, we can phrase analogous origin-destination questions but with no need to aggregate to areal units in order to consider them.  Rather, we build a \emph{joint}  intensity of the form $\lambda(\bm{s}_{o}, \bm{s}_{d})$ over pairs of locations $(\bm{s}_{o}, \bm{s}_{d}) \in D_{o} \times D_{d}$ where $\bm{s}_{o}$ is a theft location within a region $D_{o}$ and $\bm{s}_{d}$ is a recovery location within a region $D_{d}$.  In our application, the origin region and the destination region are the same, i.e., $D_{o} = D_{d} = D$.

To learn about the flow of vehicles from theft location to recovery location, suppose we choose a subregion, say $B_{o} \subset D_{o}$, as a theft neighborhood and we choose a subregion, $B_{d} \subset D_{d}$, as a recovery neighborhood.  Then, working with Poisson processes, as below, integrating an intensity over a subregion provides the distribution of the number of events in the region.  Therefore, we can investigate the incidence (or chance) of recovery in $B_{d}$ with theft in $B_{o}$.

Examination of modeling of spatial interaction has a long history in the literature. \cite{Wilson(75)} provides an early review.
\cite{Fotheringham(83)} presents a more formal discussion. More recent reviews can be found in \cite{RoyThill(03)} and in \cite{LeSagePace(08)}.
Spatial interaction data have become increasingly available due to the wide adoption of location-aware technologies \citep{Guoetal(12)}.
Examination of mobility data also has some history, e.g., \cite{BrownHolmes(71)}, \cite{Simpson(92)} and more recently, \cite{deVriesetal(09)}.
Origin-destination problems involving mobility can be found in, e.g., \cite{Woodetal(10)}; \cite{AdrienkoAdrienko(11)}; \cite{Guoetal(12)}.

For us, mobility refers to the movement of a vehicle from a theft location to a recovery location.
Pertinent to our setting is work of \cite{AssuncaoLopes(07)} and \cite{LopesAssuncao(12)}.  Particularly, the former builds a bivariate linked point process with a joint pairwise interaction function.  Our view is that the theft locations should be viewed as conditionally independent given the intensity function so that either a nonhomogeneous Poisson process (NHPP) or a log-Gaussian Cox process (LGCP) model applies.


There is little discussion connecting origin-destination problems and spatial point processes in the literature. \cite{Benesetal(05)} consider statistical analysis of linked point processes, where, in their study, for each case of a disease they have the coordinates of the individual's home and of the reported infection location. However, they used only the distance between the two linked locations.
Again, \cite{AssuncaoLopes(07)} and \cite{LopesAssuncao(12)} consider bivariate linked point processes as point processes with events marked with another spatial event representing origin-destination data types. Their methods are illustrated with the Belo Horizonte data on car theft locations and the eventual car retrieval locations; this data is also analyzed here.

As noted above, three types of issues are considered with regard to automobile theft data and we devote a section below to each. First, the set of theft locations is modeled by using both a nonhomogeneous Poisson process as well as a log-Gaussian Cox process.  Second, a conditional regression specification is proposed to provide the distribution of recovery location given theft location.
Third, we investigate the dependence between the theft location point pattern and the recovery location point pattern.  We consider a joint model, viewing the data as an origin-destination pair of points, and treating the point pattern data as consisting of random pairs of locations.  Because both origin and destination are points in two-dimensional space, $\mathbb{R}^{2}$, we specify the model as a point pattern over a bounded set $D_{o} \times D_{d}\subset \mathbb{R}^{2}\times \mathbb{R}^{2}$. This approach needs an intensity over $D_{o} \times D_{d}$ linking pairs of locations and introduces two Gaussian processes.  We note that, in the model fitting using Markov chain Monte Carlo, we use elliptical slice sampling  (\cite{MurrayAdams(10)}, \cite{MurrayAdamsMacKay(10)}, \cite{Leininger(14)}) whenever Gaussian processes are specified.  Elliptical slice sampling provides an efficient, tuning-free Gaussian process sampler.

An important last point here is that we do not attempt to propose a common model to address these three issues.  Each issue suggests its own probabilistic specification.  In particular, the first issue has no need to model recoveries, the second issue has no need to model thefts, and the third issue does not yield a conditional geostatistical regression model for recoveries given thefts.

One could view the model in Section 4 as a submodel of that in Section 5 if we propose a specification which thinks of the recovery locations as bivariate marks associated with the theft locations. That is, we can offer a single point pattern model for theft locations and then a ``response'' model for the location of the recovery associated with that theft location; the recovery locations are not viewed as a point pattern.  However, our version seeks to think about observing two point patterns which are dependent, encouraging joint bivariate modeling of them.

The plan of the paper is the following. Section 2 provides a description of the two datasets that motivate this paper. Section 3 presents the statistical approach that models the set of theft locations using both a nonhomogeneous Poisson process as well as a log-Gaussian Cox process. Then Section 4 considers the conditional specification approach which provides the distribution of recovery locations given a fixed set of theft locations. Section 5 supplies a joint modeling approach, viewing the data as an origin-destination pair, and treating the point pattern as consisting of random pairs of locations.  The paper ends with a summary and future work.

\section{Data Description}
\label{sec:Data}

We analyze two datasets consisting of a collection of automobile thefts and recoveries, one for the state of Neza in Mexico, the other in Belo Horizonte in Brazil.
The supplied longitude and latitude information is transformed to eastings and northings on km scale in Figures 1-3 below.
There may be measurement error in the locations but this is beyond the scope of the data we have to work with.

\subsection{The Neza data}
\label{sec:Neza}

The Ciudad Neza (referred to as Neza in what follows) is a city and municipality adjacent to the northeast corner of Mexico's Federal District.
It is part of the Mexico City metropolitan area.
The region consists of a North and a South part, separated by a single road.
On the east side of this road there is a large park and on the west side an airport.
In the analysis below, the North and South parts are separated.
Our dataset contains car theft locations in 2015.
The number of car theft locations is 4,016, after deleting some missing locations.

We also have several covariates, available at areal unit scale and split into two categories.
The first category consists of population types and is available through the Town Hall of Neza as well as
the Department of Citizen Security in Neza: (1) \texttt{Pop15} - number of individuals 15 years and older, (2) \texttt{Apart} - number of apartments, (3) \texttt{Eco} - number of economically active individuals, (4) \texttt{Employ} - number of employed individuals, hence \texttt{unEmploy} - number of unemployed individuals, (5) \texttt{inBorn} - number of individuals born in the area, hence \texttt{outBorn} - number of individuals born outside the area, (6) \texttt{Health} - number of individuals with health insurance access, hence \texttt{noHealth} - number of individuals without health insurance access and (7) \texttt{Scholar} average of scholarly grade (integer values from $6- 10$) which reflects the number of years of received education for citizens in an areal unit (on average and then discretized to the closest integer).

\begin{figure}[htbp]
 \begin{minipage}{0.48\hsize}
  \begin{center}
   \includegraphics[width=7.5cm]{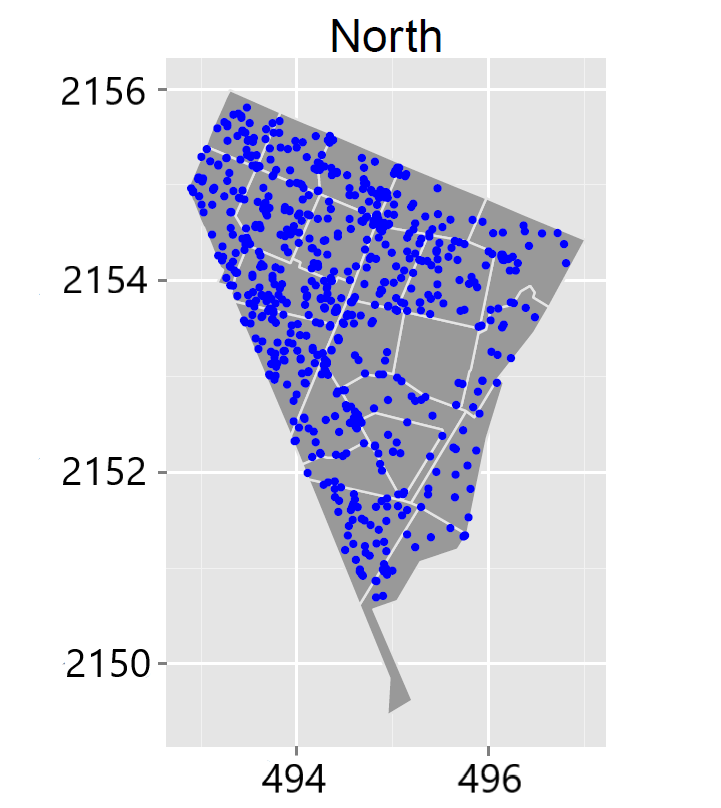}
  \end{center}
 \end{minipage}
 \hfill
 \begin{minipage}{0.48\hsize}
  \begin{center}
   \includegraphics[width=7.5cm]{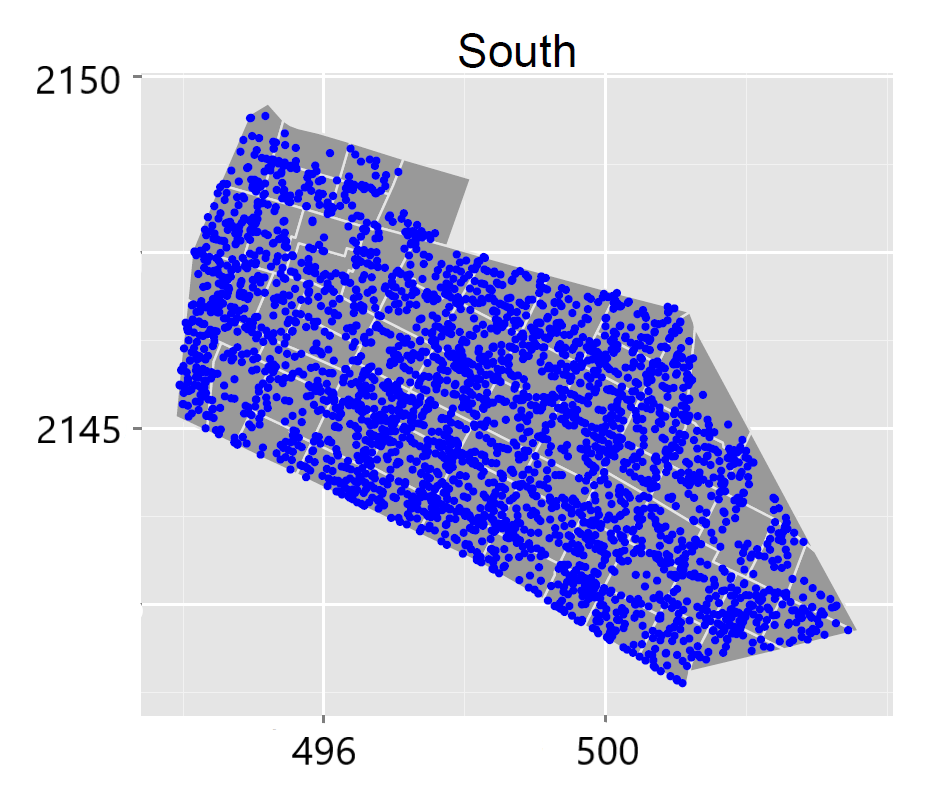}
  \end{center}
 \end{minipage}
  \caption{Car theft locations in the North region (left) and the South region (right) in Neza ($x$-axis (easting) and $y$-axis (northing) are at km scale).}
  \label{fig:Map}
\end{figure}

The second category consists of crime types, available from the Department of Citizen Security in Neza: (1) \texttt{Extor} - number of extortion crimes, (2) \texttt{Murder} - number of murders, (3) \texttt{Burg} - number of burglaries, (4) \texttt{Shop} - number of shop robberies, (5) \texttt{Public} - number of public transport robberies, (6) \texttt{Street} - number of street robberies, (7) \texttt{Kidnap} - number of kidnappings and (8) \texttt{Total} - total number of infractions (some additional crimes beyond (1) through (7) are included here).
These covariates are provided for 90 disjoint areal units, referred to as blocks, in Neza.
Figure \ref{fig:Map} shows the theft locations for the North and South regions.
22 blocks are located in the North region with the remaining 68 blocks in the South region.  They are shown in white in the figure.  Of the thefts,
3,327 points (689 points) are observed in the South (North) region.  These locations do not seem to suggest concentration in ``hot spots'' \citep{Weisburd(15)}.

Unfortunately, the recovery locations are observed for only 382 of the theft locations.
This is a commonly encountered situation in the context of car theft and recoveries \citep[see, e.g.,][]{Report(98)}.  Based on \cite{Report(98)}, the number of recoveries moved around 12$\%$ in 1998 in US. Neza is moving around this figure although fifteen years after that report.
The set of recovery locations can be relatively small compared to the set of theft locations.
In particular, the structure of the Mexican police forces is such that local forces have authority over their region, perhaps their city.  Information from other police departments is sometimes not accessible. For us, this means that the Neza police have authority only over the recoveries within the Neza region. That is why we only have data within Neza and may help to explain the low recovery rate.

\begin{figure}[htbp]
 \begin{minipage}{0.48\hsize}
  \begin{center}
   \includegraphics[width=8cm]{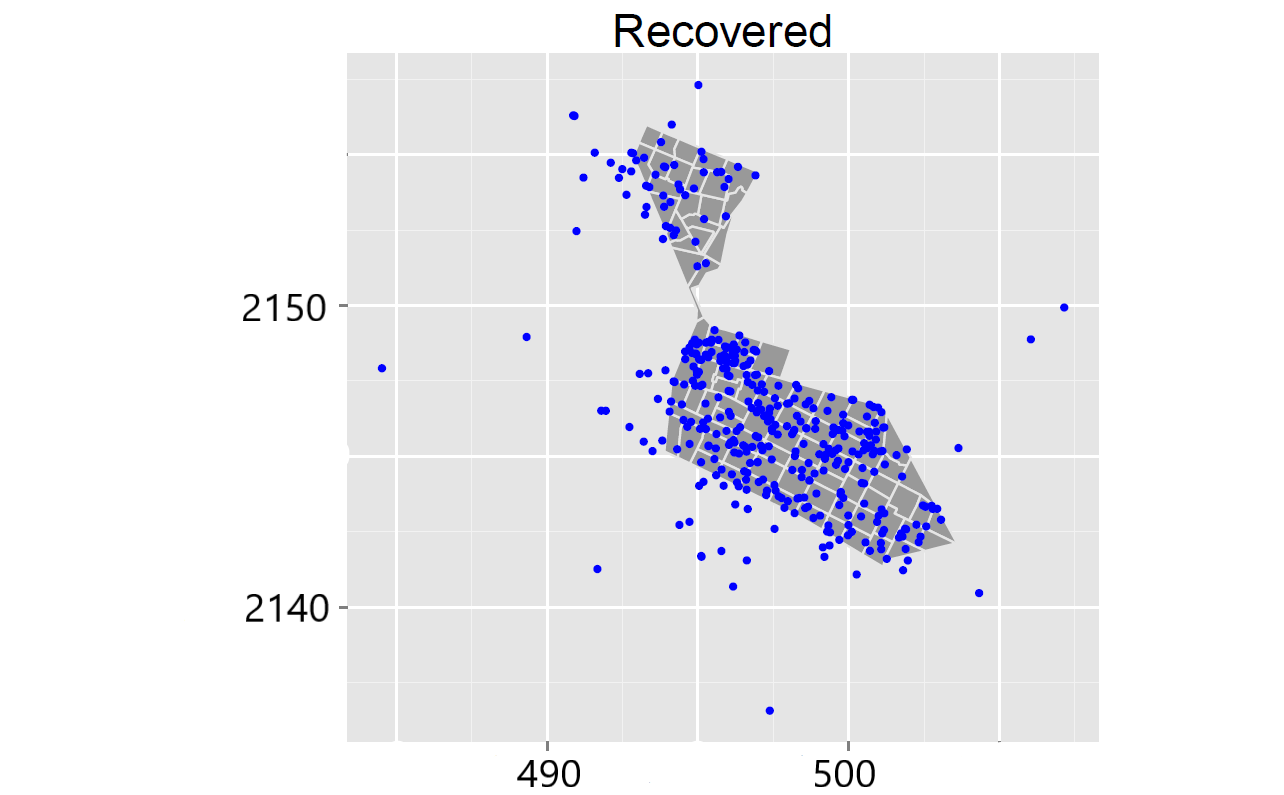}
  \end{center}
 \end{minipage}
 \hfill
 \begin{minipage}{0.48\hsize}
  \begin{center}
   \includegraphics[width=8cm]{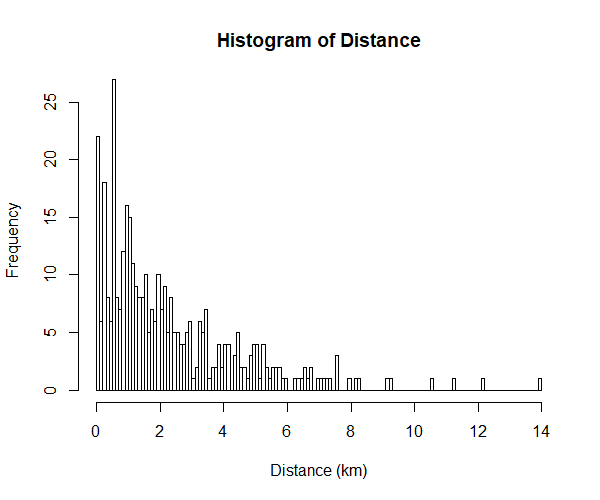}
  \end{center}
 \end{minipage}
  \caption{Recovery locations (left, $x$-axis (easting) and $y$-axis (northing) are at km scale) and histogram of the distance between theft and recovery locations (right, $x$-axis (distance) also at km scale).}
  \label{fig:Pairs}
\end{figure}

Figure \ref{fig:Pairs} shows the plot of the recovery locations for the observed theft and recovery pairs as well as a histogram of the distance between theft and recovery locations. Looking at Figure \ref{fig:Pairs}, we see that the recoveries tend to be close to the theft location. This closeness between recovery location and theft location motivates our ensuing modeling strategies. This closeness was also found in \cite{Lu(03)} and \cite{SureshTewksbury(13)}. Other than locations, we have no individual vehicle data. However, the police published an internal report in which they described the cars most often stolen in Neza during the period 2013-2016. Pick-up trucks dominate the list. The police confirmed that the black market was not the main aim of a car theft; perhaps the theft was made to move (stolen) goods from one place to another. This could explain why the recovery locations within the region tend to be close to the theft locations.  We note this is in accord with \cite{Morganetal(16)} who observe that the majority of recovered vehicles occur because the theft was motivated by joy-riding, by the need to travel, or by desire for the contents of the vehicle rather than the vehicle itself.

\subsection{The Belo Horizonte data}
\label{sec:Belo}

\begin{figure}[htbp]
 \begin{minipage}{0.48\hsize}
  \begin{center}
   \includegraphics[width=8cm]{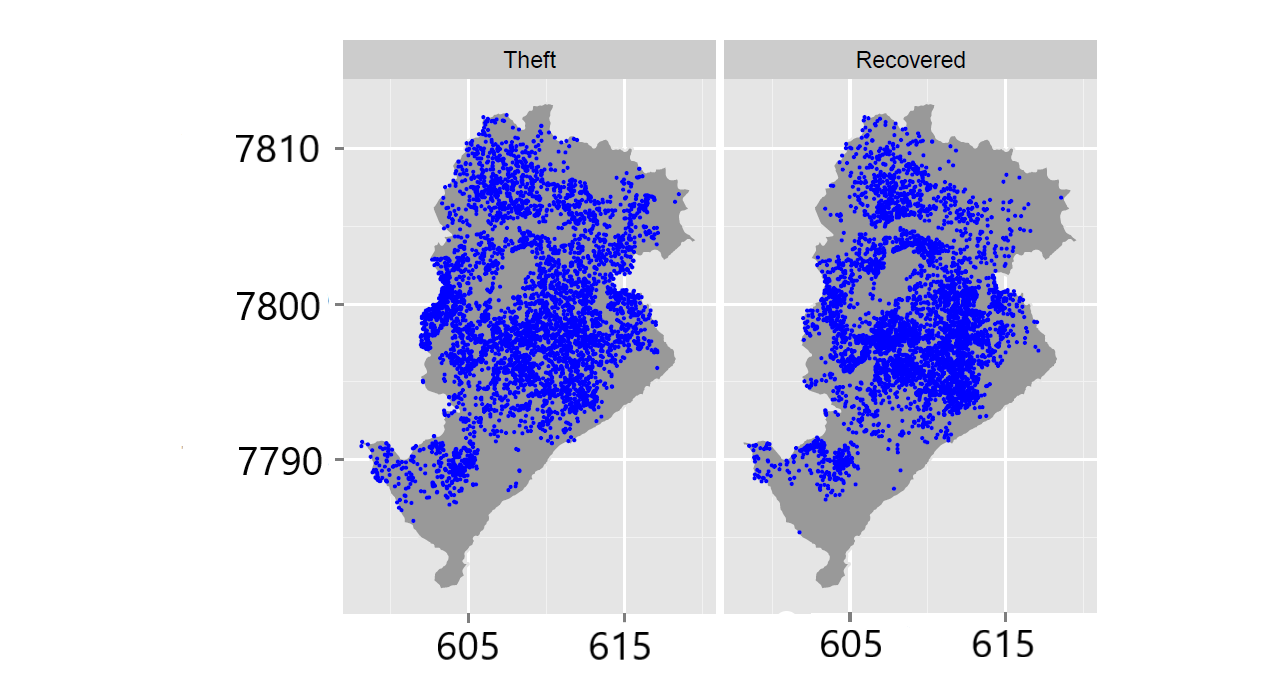}
  \end{center}
 \end{minipage}
 \hfill
 \begin{minipage}{0.48\hsize}
  \begin{center}
   \includegraphics[width=8cm]{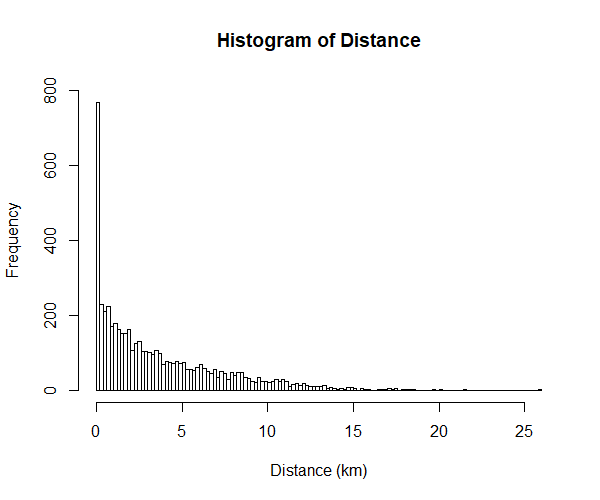}
  \end{center}
 \end{minipage}
  \caption{Car theft and recovery locations (left, $x$-axis (easting) and $y$-axis (northing) are at km scale) and histogram of the distance between theft and recovery locations (right, $x$-axis (distance) also at km scale) in Belo Horizonte}
  \label{fig:MapBH}
\end{figure}

We also examine car theft and recovery point patterns in Belo Horizonte in Brazil (\cite{AssuncaoLopes(07)}).
The dataset contains 6,339 thefts during the study window with 5,250 eventually found within the city limits. So, there is a much higher recovery rate for this data than for the Neza data. (Belo Horizonte is a large city with a population of nearly 1.5 million people and much more extensive police coverage, in part explaining the higher recovery rate.)  Furthermore, we only received the theft locations for the cars that were recovered. Additionally, this dataset does not have any covariate information. It may be argued that there is potential bias in this subsample of thefts. This cannot be assessed but with nearly $85\%$ of the total thefts included, we hope that the bias is small.
The left panel of Figure \ref{fig:MapBH} shows the point patterns of theft and recovery locations.
The point patterns are similar, though recovery points seemed to be a bit more concentrated. The right panel provides the histogram of the distance between theft and recovery locations.  Again, recovery location tends to be near theft location; in fact, 770 pairs (roughly $15\%$) are observed to be within 200m of each other.

\section{Modeling of car thefts}
\label{sec:Model}

\subsection{LGCP and NHPP models for vehicle theft}
\label{sec:LGCPNHPP}

Here we turn to the first issue raised in the Introduction.  Viewing the collection of car thefts as a random point pattern, can a satisfying explanatory model be developed?  We seek to provide an investigator with understanding of the nature of the intensity surface that is driving the point pattern of thefts.  This surface can be viewed as a risk surface for theft, enabling clarification of where risk is high, where it is low.

To develop a rich intensity surface, we only consider the vehicle theft events in Neza, employing the available covariate information.
Let $\mathcal{S}=\{\bm{s}_{1},\ldots,\bm{s}_{n}\}$ denote the observed point pattern over the study region $D\subset \mathbb{R}^{2}$.  In our case, $\mathcal{S}$ denotes the set of car theft locations and $D$ is either the North region or the South region.  We fit the Neza data for the North and South subsets separately due to the fact that there is only one road between the North and the South.  It is awkward to work with a common LGCP intensity over two areas which are essentially disjoint; it is more comfortable to fit a separate intensity to each.
We view the theft events as conditionally independent given the intensity and therefore consider a non-homogeneous Poisson process (NHPP) and a log-Gaussian Cox processes (LGCP, \cite{Molleretal(98)}) for modeling theft events.

The LGCP is defined so that the log of the intensity is a Gaussian process (GP), i.e.,
\begin{align}
\log \lambda(\bm{s})=\bm{X}(\bm{s})\bm{\beta}+z(\bm{s}), \quad  \bm{z}(\mathcal{S})\sim \mathcal{N}(\bm{0}, \mathbf{C}_{\bm{z}}), \quad \bm{s}\in D.
\end{align}
where $\bm{X}(\bm{s})$ is a covariate vector at $\bm{s}$ and $z(\bm{s})$ is a Gaussian process.
In particular, the point pattern $\mathcal{S}$ has associated vector $\bm{z}(\mathcal{S})=(z(\bm{s}_{1}), \ldots, z(\bm{s}_{n}))$ which follows an $n$-variate zero mean Gaussian distribution, with covariance matrix $\mathbf{C}_{\bm{z}}=[C(\bm{s}_{i}, \bm{s}_{j})]_{i,j=1,\ldots, n}$.  The component spatial random effects for the intensity surface provide local pushing up and pulling down the surface, as appropriate. We assume an exponential covariance function, i.e., $C(\bm{u}, \bm{u}^{'})=\sigma^2 \exp(-\phi \|\bm{u}-\bm{u}^{'}\|)$\footnote{Since we never see observations of the intensity, it is hard to justify or identify a richer covariance function for the spatial random effects.}.

If $\bm{z}(\bm{s})$ is removed from the log intensity, the corresponding NHPP is obtained. NHPP's have a long history in the literature (see, e.g., \cite{Illianetal(08)}).
Furthermore, given $\lambda(\bm{s})$ with $z(\bm{s})$ included, $\mathcal{S}$ again, follows an NHPP with intensity $\lambda(\bm{s})$.
The likelihood takes the form
\begin{align}
\mathcal{L}(\mathcal{S})&\propto \exp\biggl(-\int_{D}\lambda(\bm{u})d\bm{u} \biggl)\prod_{i=1}^{n}\lambda(\bm{s}_{i})
\end{align}

For inference with a LGCP using (2), the stochastic integral inside the exponential need to be approximated. We create $K$ grid cells roughly uniformly over the study region $D$; convergence to the exact posterior distribution when $K \to \infty$ (with grid cell area decreasing to $0$) is guaranteed following \cite{Waagepetersen(04)}.    Then, the approximate likelihood for the LGCP becomes
\begin{align}
\mathcal{L}(\mathcal{S})\propto \exp\biggl(-\sum_{k=1}^{K}\lambda(\bm{u}_{k})\Delta_{k} \biggl)\prod_{k=1}^{K}\lambda(\bm{u}_{k})^{n_{k}}
\end{align}
where $n_{k}$ is the number of points in the $k$-th grid, i.e., $\sum_{k}^{K}n_{k}=n$, $\Delta_{k}$ is the area of $k$-th grid (in practice, we standardize $\Delta_{k}$ so that $\sum_{k}^{K}\Delta_{k}=|D|=1$) and $u_{k}$ is the ``representative point" for the $k$-th grid (e.g., \cite{MollerWaagepetersen(04)} and \cite{BanerjeeCarlinGelfand(14)}).  In fact, since covariate values for 90 different areal units are available, we adopt this as our discretization.  In order to supply full inference we work within a Bayesian framework, fitting the model using Markov chain Monte Carlo (see, e.g., \cite{RobertCasella(04)}).
Other model fitting approaches are available which may be helpful for large point patterns requiring a large number of grid cells.  They include integrated nested Laplace approximation (INLA) \citep{Simpsonetal(16b)}, approximate Gaussian process models, e.g, nearest neighbor Gaussian processes \citep{Dattaetal(16a)} and multi-resolution Gaussian processes \citep{Katzfuss(17)}.

\subsection{Covariate selection}
\label{sec:Covariate}

With regard to covariate selection, the spatstat R-package (\cite{BaddeleyTurner(05)}; \cite{Baddeleyetal(13)}) supports the model fitting of spatial point processes, in particular Poisson processes, and related inference and diagnostic tools.
The function \texttt{ppm} fits a spatial point process to an observed point pattern and allows the inclusion of covariates.

We note that implementing fully Bayesian variable selection in this setting of log Gaussian Cox processes brings us to new territory which has not been studied and is beyond our scope here.  We introduce naive variable selection solely to obtain a more manageable number of variables for the sizes of the point patterns. We should view our suggested significances as somewhat ad hoc, and our inference as illustrative.

Working with the 90 blocks (22 in North and 68 in South), covariates from the 15 listed in Section 2.1 are chosen by forward and backward selection (\texttt{step} function in R) based on the models fitted by the \texttt{ppm} function\footnote{Using just forward or just backward selection produced the same selection.}.
%

With the NHPP model, the forward and backward algorithm is implemented (\texttt{step} function) using the Bayesian Information Criterion (BIC) penalty ($\log(n)$) for each region.  What emerged, for the South and North, respectively, is
\begin{align}
\bm{X}_{S,k}\bm{\beta}&=\beta_{0}+\beta_{1}\texttt{Extor}_{k}+\beta_{2}\texttt{Shop}_{k}  +\beta_{3}\texttt{Street}_{k}+\beta_{4}\texttt{Total}_{k} \nonumber \\
&+\beta_{5}\texttt{Eco}_{k}+\beta_{6}\texttt{Scholar}_{k} \\
\bm{X}_{N,k}\bm{\beta}&=\beta_{0}+\beta_{2}\texttt{Shop}_{k}+\beta_{4}\texttt{Total}_{k}+\beta_{7}\texttt{Apart}_{k}.
\end{align}
All covariates are centered and scaled. The same covariates are used for the LGCP model.

Using these forms in Bayesian model fitting, Markov chain Monte Carlo is implemented.  For sampling of $\bm{\beta}$ in the LGCP and NHPP, an adaptive random walk Metropolis-Hastings (MH) algorithm (\cite{AndrieuThoms(08)}) is employed.
Elliptical slice sampling for the Gaussian processes in the LGCP (\cite{MurrayAdams(10)}, \cite{MurrayAdamsMacKay(10)}, \cite{Leininger(14)}); it is an efficient, tuning-free Gaussian process sampler.
20,000 samples are discarded as the burn-in period and a subsequent 20,000 samples are preserved as posterior samples for the LGCP and for the NHPP, respectively.
Since, for spatial Gaussian processes, $\phi$ and $\sigma^2$ are not identifiable but the product, $\phi \sigma^2$ is  \citep{Zhang(04)}, an informative prior distribution needs to be adopted for one of them. Here, we assume informative support for $\phi$ and adopt an inverse Gamma distribution for $\sigma^2$ with relatively large variance.  As specific prior settings, we assume $\sigma^2\sim \mathcal{IG}(2,0.1)$ (inverse gamma), $\bm{\beta}\sim \mathcal{N}(\bm{0}, 100\mathbf{I})$ (normal) and $\phi \sim \mathcal{U}[0, 10]$ (uniform), where, after rescaling, as above, the easting and northing distances are in kilometers.

For the South region, all coefficients were significant, i.e., a $95\%$ credible interval (CI) doesn't include 0, under the NHPP while $\beta_{5}$ was insignificant under the LGCP. For the North region, again all coefficients were significant under the NHPP while $\beta_{7}$ was insignificant under the LGCP.  Details are omitted but we note that the total number of infractions has a large positive (increasing) effect on theft events for both the North and South regions.
For the South region, the posterior log likelihood for the NHPP can be summarized (posterior mean, $95\%$ credible interval) as $-294.3 \hspace{.2cm} (-298.6, -291.7)$ while that summary for the LGCP is $-228.2 \hspace{.2cm} (-242.4, -216.7)$.   For the North region, the posterior log likelihood for the NHPP can be summarized as $-81.98 \hspace{.2cm} (-.85.44, -80.28)$ while that summary for the LGCP is $-68.47 \hspace{.2cm} (-75.46, -62.68)$.  The larger likelihood shows that the LGCP emerges as preferred for both regions.  We provide further support for the LGCP through cross-validation in the next subsection.

\subsection{$p$-thinning cross validation}
\label{sec:CV}
One of standard approaches for assessing model adequacy and is cross validation which is available for point pattern models with conditionally independent locations given the intensity \citep[see,][]{LeiningerGelfand(17),ShirotaGelfand(17)}. The ensuing discussion in this subsection follows \cite{LeiningerGelfand(17)} and \cite{ShirotaGelfand(17)}.

Cross validation is implemented by splitting the whole dataset into a training and a testing dataset using $p$-thinning as proposed by \cite{LeiningerGelfand(17)}.
Let $p$ denote the preservation probability, i.e., we keep $\bm{s}_{i}\in \mathcal{S}$ with probability $p$ and delete $\bm{s}_{i}\in \mathcal{S}$ with probability $1-p$.
This generates a training point pattern $\mathcal{S}^{train}$ and a test point pattern $\mathcal{S}^{test}$, which are independent, conditional on intensity $\lambda(\bm{s})$.
Especially, $\mathcal{S}^{train}$ has intensity $\lambda(\bm{s})^{train}=p \lambda(\bm{s})$. We set $p=0.5$ and estimate $\lambda(\bm{s})^{train}$ $\bm{s}\in D$.  Then, the posterior samples of $\lambda^{train}(\bm{s})$ are converted into predictive samples of $\lambda^{test}(\bm{s})$ using $\lambda^{test}(\bm{s})=\frac{1-p}{p}\lambda^{train}(\bm{s})= \lambda^{train}(\bm{s})$.

Let $\{B_{r}\}$ be a collection of subsets of $D$ as an evaluation grid. For the choice of  $\{B_{r}\}$, \cite{LeiningerGelfand(17)} suggest to sample random subsets of the same size uniformly over $D$ (not necessarily disjoint). Specifically, for $q \in (0,1)$, if the area of  each $B_{r}$ is $q|D|$, then $q$ is the \emph{relative} size of each $B_{r}$ to $D$.
Based on the $p$-thinning cross validation, two model performance criteria are considered: (i) predictive interval coverage (PIC) and  (ii) rank probability score (RPS). PIC offers assessment of model adequacy, RPS enables model comparison.
\\\\
\noindent{\bf Predictive Interval Coverage}\\

After the model fitting to $\mathcal{S}^{train}$, the posterior predictive intensity function can provide posterior predictive point patterns and therefore samples from the posterior predictive distribution of $N(B_{r})$ for each $r$. For the $\ell$-th posterior sample, $\ell = 1,...., L$, the associated predictive residual is defined as
\begin{align}
R_{\ell}^{pred}(B_{r})=N^{test}(B_{r})-N^{(\ell)}(B_{r})
\end{align}
where $N^{test}(B_{r})$ is the number of points of the test data in $B_{r}$ and $N^{(\ell)}(B_{r})$ is the  number of predictive points in $B_{r}$, simulated under the parameters generated from $\ell$-th MCMC iteration.
If the model is appropriate, the empirical predictive interval coverage rate, i.e., the proportion of intervals which contain $0$, is expected to be the nominal level of coverage (below, $90\%$ nominal coverage is chosen). Empirical coverage much less than the nominal suggests model inadequacy; predictive intervals are too optimistic. Empirical coverage much above, for example $100\%$, is also undesirable. This indicates that more uncertainty is introduced into the model than needed.
\\\\

\noindent{\bf Rank Probability Score}\\

\cite{GneitingRaftery(07)} propose the continuous rank probability score (CRPS).
This score is derived as a proper scoring rule and a criterion  for evaluating the precision of a predictive distribution for continuous variables.
In our context, we use a rank probability score for count data, which is discussed in \cite{Czadoetal(09)} and references therein.
Intuitively, a good model will provide a predictive distribution that is concentrated around test counts.
Although the RPS has a complex formal computational form, it can be directly calculated by Monte Carlo integration.
In particular, for a given $B_{r}$, the RPS is calculated as
\begin{align}
\text{RPS}(F, N^{test}(B_{r}))&=\frac{1}{L}\sum_{\ell=1}^{L}|N^{(\ell)}(B_{r})-N^{test}(B_{r})| \nonumber \\
&-\frac{1}{2L^2}\sum_{\ell=1}^{L}\sum_{\ell^{'}=1}^{L}|N^{(\ell)}(B_{r})-N^{(\ell^{'})}(B_{r})|
\end{align}
Summing over the collection of $B_{r}$ gives a model comparison criterion.
Smaller values of the sum are preferred.

\begin{figure}[htbp]
 \begin{minipage}{0.48\hsize}
  \begin{center}
   \includegraphics[width=7cm]{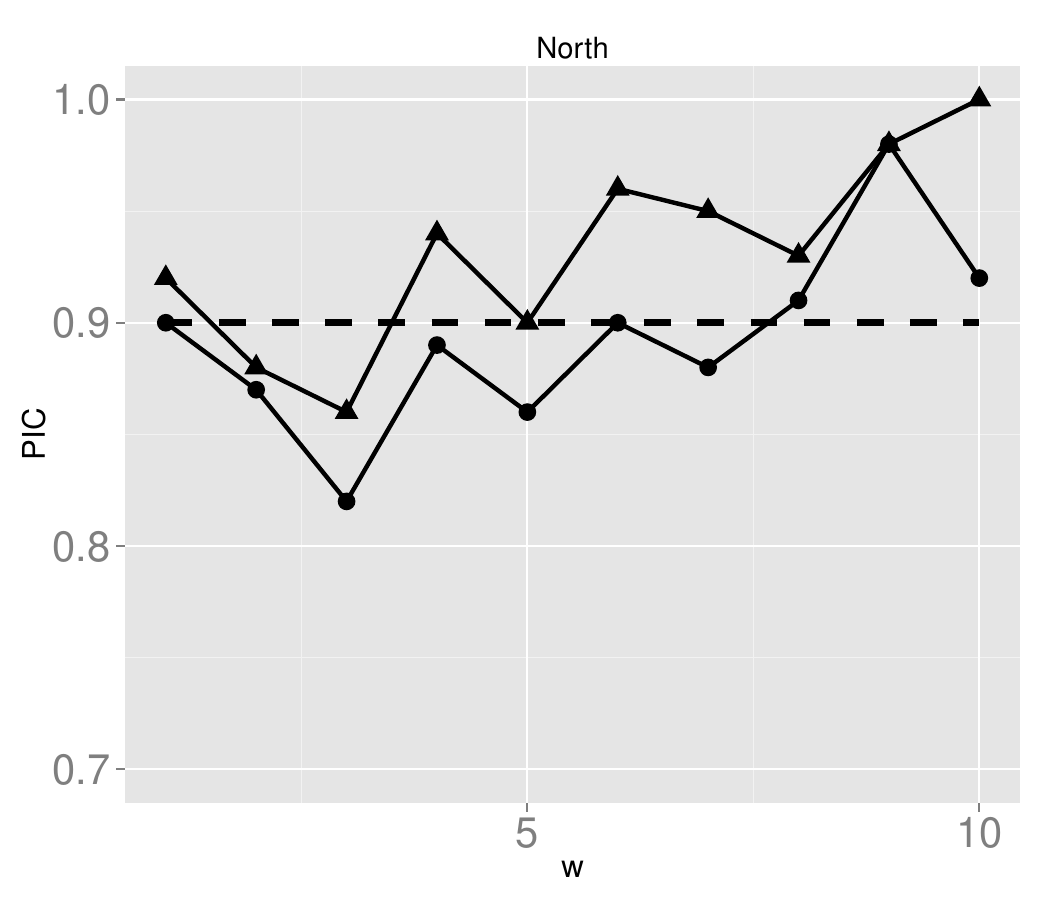}
  \end{center}
 \end{minipage}
 \hfill
 \begin{minipage}{0.48\hsize}
  \begin{center}
   \includegraphics[width=7cm]{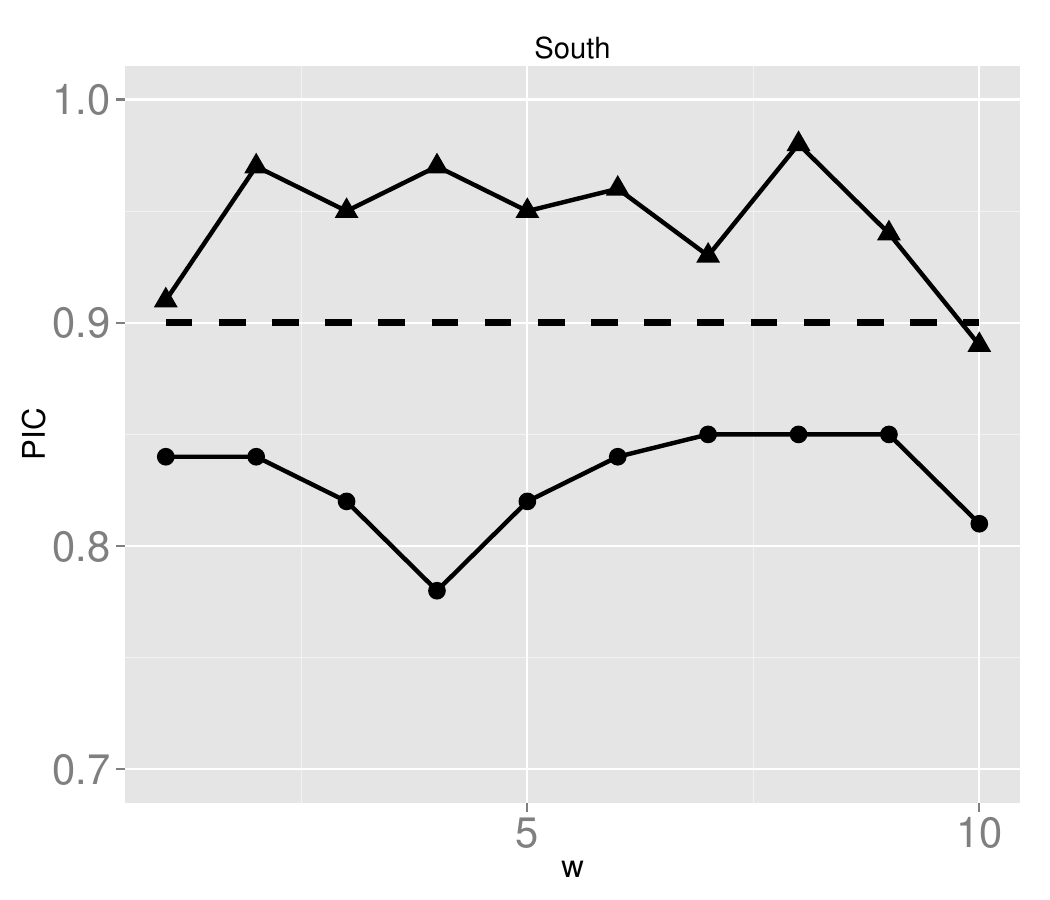}
  \end{center}
 \end{minipage}
 \begin{minipage}{0.48\hsize}
  \begin{center}
   \includegraphics[width=7cm]{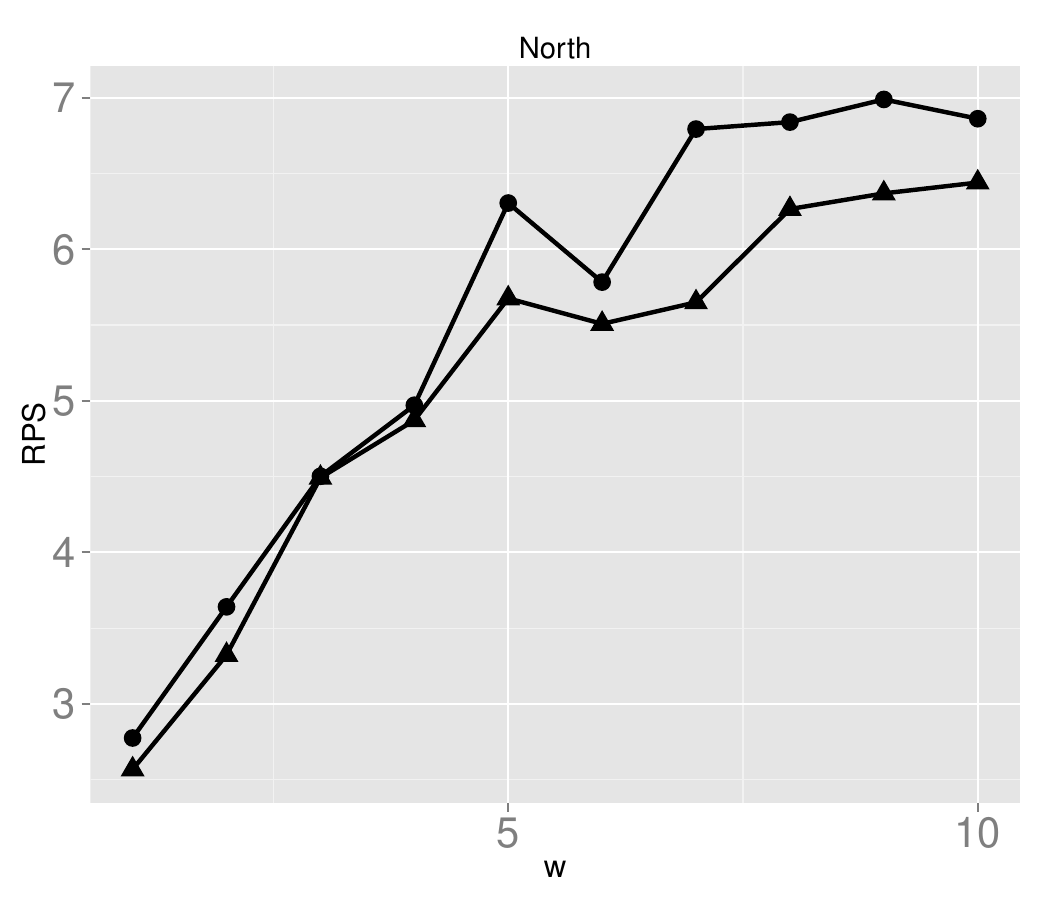}
  \end{center}
 \end{minipage}
 \hfill
 \begin{minipage}{0.48\hsize}
  \begin{center}
   \includegraphics[width=7cm]{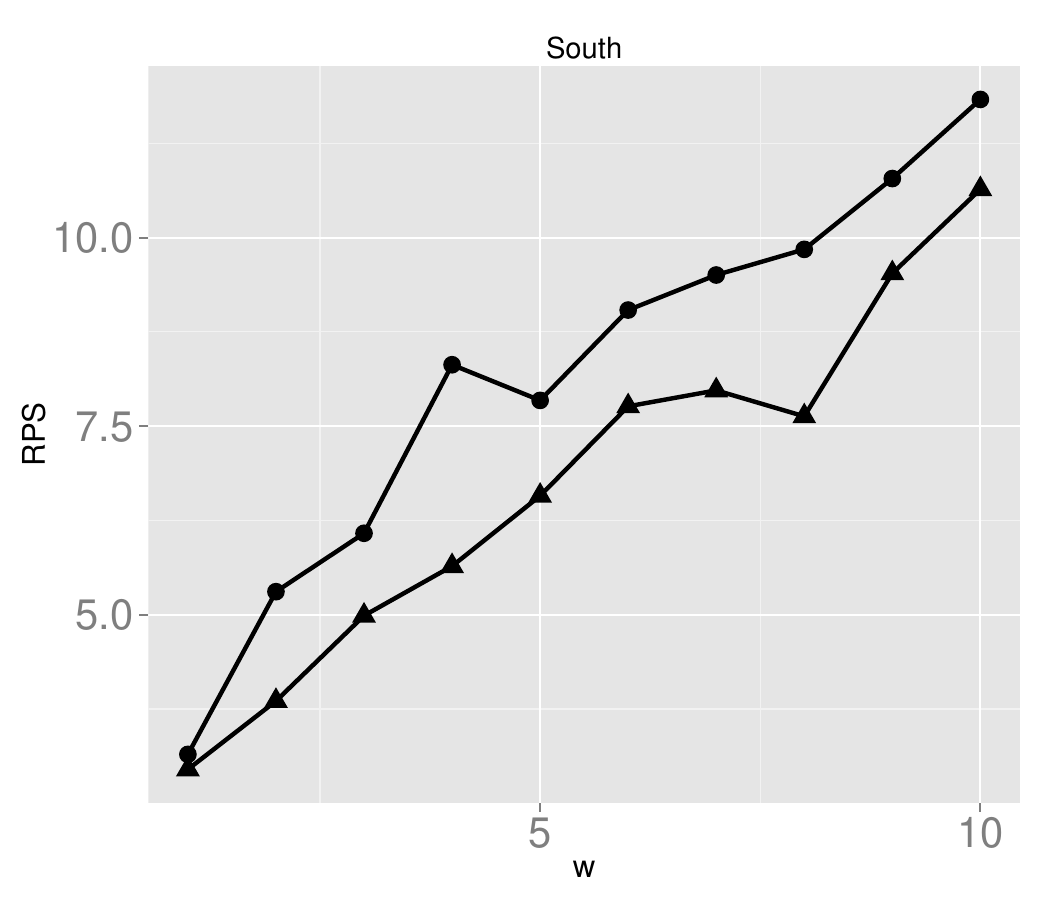}
  \end{center}
 \end{minipage}
  \caption{PIC (top) with $90\%$ nominal level (dashed line) and RPS (bottom) for the North (left) and South (right) regions: NHPP ($\bullet$) and LGCP ($\blacktriangle$). $w$ is the number of randomly selected blocks for model comparison.}
  \label{fig:MVTheft}
\end{figure}

The results of model validation are presented for the Neza data using predictive interval coverage and ranked probability score.
We set $p=0.5$ for dividing into training and test datasets.
Figure \ref{fig:MVTheft} shows the PIC with 90$\%$ nominal level and the RPS for both regions.
Here, $w$ denotes the number of randomly selected blocks for model comparison.
As for the choice of ${B_{r}}$, since the total number of grid cells for this dataset is small, here we choose $w=1,\ldots, 10$ grids from the 22 grids in the North and 68 grids in the South, rather than choosing ${B_{r}}$ with respect to a rate $q$.
Again, the LGCP outperforms the NHPP, more so for the South, the larger dataset.

\begin{figure}[htbp]
 \begin{minipage}{0.48\hsize}
  \begin{center}
   \includegraphics[width=8cm]{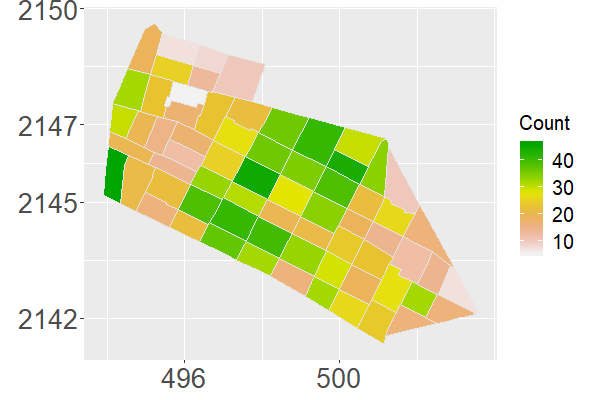}
  \end{center}
 \end{minipage}
 \hfill
 \begin{minipage}{0.48\hsize}
  \begin{center}
   \includegraphics[width=8cm]{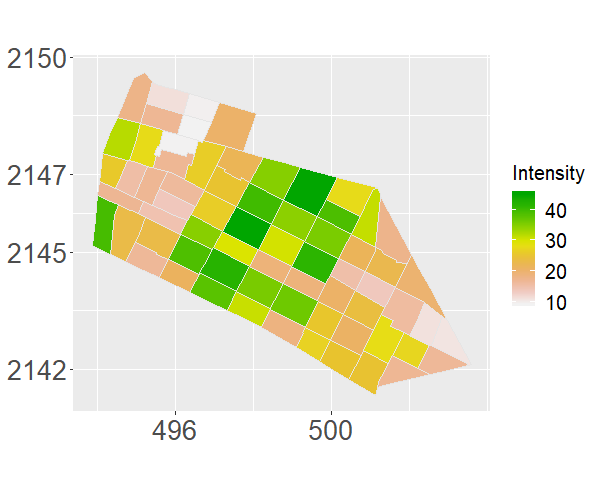}
  \end{center}
 \end{minipage}
  \caption{Testing data counts (left) and posterior predictive intensity surface, LGCP (right) in the South region.}
  \label{fig:LGCPsouth}
\end{figure}

\begin{figure}[htbp]
 \begin{minipage}{0.48\hsize}
  \begin{center}
   \includegraphics[width=8cm]{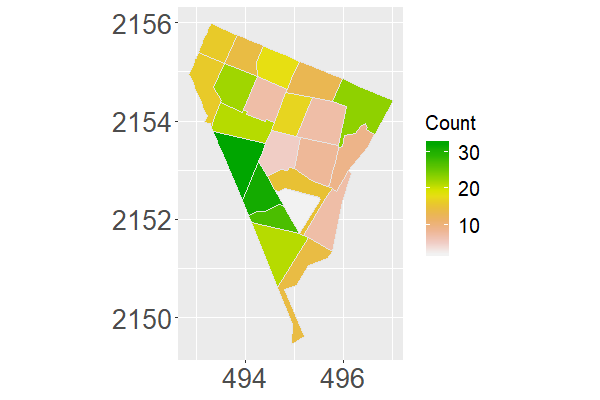}
  \end{center}
 \end{minipage}
 \hfill
 \begin{minipage}{0.48\hsize}
  \begin{center}
   \includegraphics[width=8cm]{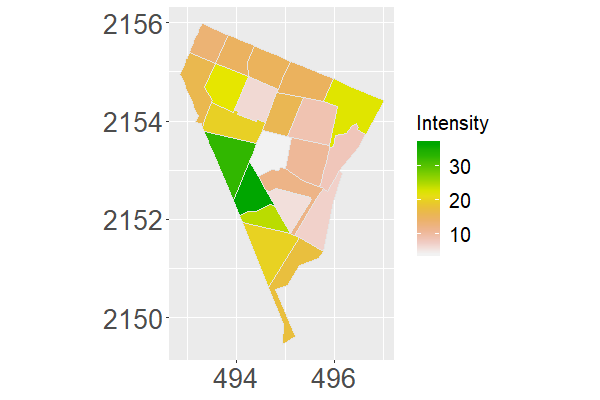}
  \end{center}
 \end{minipage}
  \caption{Testing data counts (left) and posterior predictive intensity surface, LGCP (right) in the North region.}
  \label{fig:LGCPnorth}
\end{figure}

Finally, Figures \ref{fig:LGCPsouth} and \ref{fig:LGCPnorth} display the LGCP results for the nominally $50\%$ of points in the testing sample, comparing the testing data counts with the posterior predictive intensity surface estimated by using the training data counts for the South and the North regions, respectively.
Altogether, the posterior predictive intensity surfaces well explain the distribution for the points in the testing sample for both regions.

\section{Conditioning recovery location on theft location}
\label{sec:Conditioning}

We turn to a second issue with regard to vehicle theft. How well can we predict recovery location given theft location?  Evidently, an effective predictive model would help local law enforcement in the process of vehicle recovery.  For the analysis of recovery locations, a conditional density specification given theft location is considered.
We do not have to specify a set in which our recovery locations are considered; we can include some available recovery points located outside the Neza region.
Also, we do not have to split the Neza region for this analysis.
Furthermore, we can allow the theft location to determine not only the mean for the recovery location but also the uncertainty in the recovery location.

We denote by $\bm{s}_{R}$ a recovery location and by $\bm{s}_{T}$ a theft location with the set of theft locations denoted by $\mathcal{S}_{T}=\{\bm{s}_{T,1},\ldots, \bm{s}_{T,n}\}$ and the set of recovery locations denoted by $\mathcal{S}_{R}=\{\bm{s}_{R,1},\ldots, \bm{s}_{R,m}\}$ where $m<n$.
The object of interest is the conditional density specification for recovery location $\bm{s}_{R}$ given a theft location $\bm{s}_{T}$, denoted as $f_{R}(\bm{s}_{R}|\bm{s}_{T})$.

Hence, we do not need to model $\mathcal{S}_{T}$.  Moreover, $f$ provides a distribution at each $\bm{s}_{T}$.  It is not an intensity normalized to a density for $\mathcal{S}_{R}$ over $D$.  So, $\lambda(\bm{s}_{T})f_{R}(\bm{s}_{R}|\bm{s}_{T})$ is not a version of a \emph{joint} specification for pairs of theft and recovery locations in the sense of viewing the data as a point pattern of pairs of locations over a bounded set in $\mathbb{R}^2 \times \mathbb{R}^2$.  Such a model is deferred to the next section. We could view the recovery location, when observed, as a \emph{mark} associated with the theft location but, in any event, we are imagining $f_{R}(\bm{s}_{R}|\bm{s}_{T})$ as a geostatistical specification, providing a distribution for every $\bm{s}_{R} \in D$.


Let $\mathcal{S}_{T}^{*}=\{\bm{s}_{T,1}^{*},\ldots, \bm{s}_{T,m}^{*}\}$ be the set of theft locations corresponding to recovery points, i.e., $\bm{s}_{T,j}^{*}$ is the corresponding theft location for the recovery point $\bm{s}_{R,j}$ for $j=1,\ldots, m$. For $j=1,\ldots, m$,
\begin{align}
f_{R}(\bm{s}_{R,j}|\bm{s}_{T,j}^{*})\propto |\Sigma(\bm{s}_{T,j}^{*})|^{-1/2}\exp\biggl(-(\bm{s}_{R,j}-\bm{s}_{T,j}^{*})'\Sigma(\bm{s}_{T,j}^{*})^{-1}(\bm{s}_{R,j}-\bm{s}_{T,j}^{*})\biggl), \label{eq:CD}
\end{align}
$\Sigma(\bm{s}_{T,j}^{*})$ is $2\times 2$ covariance kernel dependent on theft location $\bm{s}_{T,j}^{*}$.
A benchmark specification would assume a constant covariance kernel across theft locations, i.e.,
\begin{align}
\Sigma(\bm{s}_{T,j}^{*})=\Sigma=\begin{pmatrix}
           \sigma_{1}^2 & \rho \sigma_{1}\sigma_{2} \\
           \rho \sigma_{1}\sigma_{2} & \sigma_{2}^2
           \end{pmatrix}
\end{align}

We enrich this specification using a locally adaptive covariance kernel, employing the spatially varying covariance kernel in \cite{Higdonetal(99)},
\begin{align}
\Sigma(\bm{s}_{T})^{\frac{1}{2}}=&\sigma \begin{pmatrix}
                                      \biggl(\frac{\sqrt{4A^{2}+\|\psi(\bm{s}_{T}) \|^{4} \pi^{2}}}{2\pi}+\frac{\|\psi(\bm{s}_{T}) \|^2}{2} \biggl)^{\frac{1}{2}} & 0 \\
                                      0  & \biggl(\frac{\sqrt{4A^{2}+\|\psi(\bm{s}_{T}) \|^{4} \pi^{2}}}{2\pi}-\frac{\|\psi(\bm{s}_{T}) \|^2}{2} \biggl)^{\frac{1}{2}}
                                      \end{pmatrix} \nonumber \\
                                      &\times \begin{pmatrix}
                                      \cos(\alpha(\bm{s}_{T})) & \sin(\alpha(\bm{s}_{T})) \\
                                      -\sin(\alpha(\bm{s}_{T})) & \cos(\alpha(\bm{s}_{T}))
                                      \end{pmatrix}
\end{align}
where $\|\psi(\bm{s}_{T}) \|^2=\psi_{x}(\bm{s}_{T})^2+\psi_{y}(\bm{s}_{T})^2$, $\alpha(\bm{s}_{T})=\tan^{-1}\psi_{y}(\bm{s}_{T})/\psi_{x}(\bm{s}_{T})$ and $A=3.5$ as fixed in \cite{Higdonetal(99)}.
$\psi_{x}(\bm{s})$ and $\psi_{y}(\bm{s})$ are independent Gaussian processes with mean 0 and common Gaussian covariance function $C(\bm{s}_{T,i}, \bm{s}_{T,j})=\exp(-\phi^{*} \|\bm{s}_{T,i}-\bm{s}_{T,j}\|^2)$.
They introduce spatial dependence in $\Sigma(\bm{s}_{T})$.
$\phi^{*}$ is a tuning parameter which determines the spatial decay of the Gaussian processes. We fix this parameter at several different values in the ensuing analysis.

As a last remark here, based upon the exploratory work in Section 2 on the distance between theft location and recovery location, it seems that the bivariate normal distribution for the recovery location should be centered at the theft location.  What other location would be sensible?  Furthermore, the idea of adding regressors associated with $\bm{S}_{R}$ to explain recovery location would not be sensible; we would be regressing $\bm{s}_{R}$ on $\bm{X}(\bm{s}_{R})$. Rather, this path leads to the notion of an intensity for $\bm{s}_{R}$ over $D$ which, again, is the goal of the next section.

\subsection{Results}
\label{sec:ResultCondtional}

For recovery locations, we implement conditional density specification with constant and with spatially varying covariance kernels for both datasets.
For the constant covariance kernel parameters, we assume $\sigma_{1}^2, \sigma_{2}^2\sim \mathcal{IG}(2,0.1)$ and $\rho \in \mathcal{U}[-1,1]$.
The first 10,000 samples are discarded as burn-in and the subsequent 10,000 samples are retained as posterior samples.
For the spatially varying covariance kernel parameter, we assume $\sigma^2\sim \mathcal{IG}(2,0.1)$.
For this model, three fixed $\phi^{*}$ values are considered: (i) $\phi^{*}=30$, (ii) $\phi^{*}=10$ and (iii) $\phi^{*}=1$.
Here, the first 20,000 samples are discarded as burn-in and the subsequent 20,000 samples are retained as posterior samples.

In working with Gaussian processes, MCMC model fitting requires, at each iteration, conditional density computation for every location.  In turn, this requires repeated inversion of covariance matrices of order $n \times n$ where $n$ is the number of locations.  For the Neza dataset, the number of points is relatively small so that such inversion is manageable.  However, the number of theft locations in Belo Horizonte is much larger making MCMC model fitting very computationally demanding.
So, in this case, the study region is approximated by using 305 disjoint regular grid cells. $\Sigma(\bm{s}_{T})$ is evaluated at the nearest grid centroid.
For sampling the Gaussian processes $\psi_{x}(\bm{s})$ and $\psi_{y}(\bm{s})$, elliptical slice sampling is implemented.
The estimation results are shown in Table \ref{tab:RealCond}.
The spatially varying covariance model fits better than the constant covariance kernel model with respect to the loglikelihood, preferring the larger values of $\phi^{*}$ (weaker spatial dependence in the $\Sigma(\cdot)$'s) for the Neza, less so for Belo Horizonte.

\begin{table}[htbp]
\caption{Estimation results for the conditional model specifications}
\begin{tabular}{lcccccccc}
\hline
\hline
  &  \multicolumn{3}{c}{Neza} & & \multicolumn{3}{c}{Belo Horizonte}   \\
\hline
   & Mean & Stdev &  $95\%$ Int  & & Mean & Stdev &  $95\%$ Int   \\
\hline
 \multicolumn{2}{c}{Constant} & & & &&&\\
 $\sigma_{1}$ & 2.360 & 0.084 & [2.199, 2.542]  &  & 3.012 & 0.028 & [2.957, 3.071]  \\
 $\sigma_{2}$ & 2.142 & 0.076 & [2.001, 2.303]  &  & 3.953 & 0.038 & [3.875, 4.028] \\
 $\rho$ & -0.421 & 0.041 & [-0.502, -0.334] &   &0.039 & 0.013 & [0.013, 0.067] \\
 likelihood & -961.9 & 1.214 & [-965.0, -960.5] &  &-18269 & 1.231 & [-18272, -18268] \\
\hline
 \multicolumn{2}{c}{Spatial} & & & &&&\\
 \multicolumn{2}{c}{(i) $\phi^{*}=30$}  & & & &&&\\
 $\sigma$ & 1.527 & 0.050 & [1.430, 1.628]  && 2.796 & 0.019 & [2.757, 2.835]  \\
 likelihood & -744.2 & 15.28 & [-770.3, -711.4] &&-16632 & 16.15 & [-16666, -16603]  \\
 \multicolumn{2}{c}{(ii) $\phi^{*}=10$}  & & & &&&\\
 $\sigma$ & 1.532 & 0.049 & [1.436, 1.631] && 2.796 & 0.019 & [2.756, 2.837] \\
 likelihood & -746.6 & 14.56 & [-775.0, -717.0] &&-16631 & 16.13 & [-16660, -16598] \\
 \multicolumn{2}{c}{(iii) $\phi^{*}=1$}  & & & &&&\\
 $\sigma$ & 1.693 & 0.047 & [1.604, 1.788]  && 2.790 & 0.020 & [2.752, 2.830]  \\
 likelihood & -822.9 & 8.519 & [-838.8, -805.7]  &&-16613 & 16.93 & [-16656, -16586] \\
\hline
\hline
\end{tabular}
\label{tab:RealCond}
\end{table}


Lastly, here, model performance is compared by calculating \emph{bivariate} CRPS values, since our model is predicting a bivariate response, $\bm{s}_{R}$.   Let $\{\bm{s}_{T,h}^{test},\bm{s}_{R,h}^{test}\}_{h=1}^{H}$ be a randomly selected test sample of locations for evaluating predictive performance and $\{\bm{s}_{T,j}^{train},\bm{s}_{R,j}^{train}\}_{j=1}^{m-H}$ be the remaining training locations.
The bivariate continuous rank probability score (CRPS)  \citep{Gneitingetal(08)} values a bivariate distribution $F(\cdot|\bm{s}_{T}^{test})$ more if it is more concentrated around $\bm{s}_{R}^{test}$.  The criterion is calculated through Monte Carlo integrations, using draws from $F(\cdot|\bm{s}_{T}^{test})$, as
\begin{align}
CRPS(F(\cdot|\bm{s}_{T,h}^{test}), \bm{s}_{R,h}^{test})=\frac{1}{L}\sum_{\ell=1}^{L}\| \bm{s}_{R,h}^{(\ell)}-\bm{s}_{R,h}^{test}\|-\frac{1}{2L^{2}}\sum_{\ell=1}^{L} \sum_{\ell^{'}=1}^{L}\|\bm{s}_{R,h}^{(\ell)}- \bm{s}_{R,h}^{(\ell^{'})}. \|
\label{eq:biCRPS}
\end{align}
Here, $\bm{s}_{R,h}^{(\ell)}$ are the locations from $F(\cdot|\bm{s}_{T}^{test})$.  Fitting within a Bayesian framework enables posterior samples from $F(\cdot|\bm{s}_{T}^{test})$ using posterior samples of the model parameters.  In particular, for construction of the bivariate predictive distribution with spatially varying kernel, given the $\ell$-th posterior sample of $\psi_{x}(\bm{s})$, $\psi_{y}(\bm{s})$ at training samples $\{\bm{s}_{T,j}^{train}\}_{j=1}^{m-H}$ and $\sigma$, we generate $\psi_{x}(\bm{s})$, $\psi_{y}(\bm{s})$ at test samples $\{\bm{s}_{T,h}^{test}\}_{h=1}^{H}$ from the conditional Gaussian distribution and calculate $\Sigma(\bm{s}_{T,h}^{test})$ for $h=1,\ldots, H$. as defined in (\ref{eq:biCRPS}).

For Neza (Belo Horizonte), $H=80$ (2625) samples are randomly held out to create the testing dataset, and parameter values are estimated for the remaining 302 (2625) training dataset.
Then, given test theft locations $\{\bm{s}_{T,h}^{test}\}_{h=1}^{H}$, the predictive conditional densities are calculated for the corresponding recovery locations.  That is, for the spatially varying kernel, we predict $\psi_{x}(\bm{s}_{T,h}^{test})$ and $\psi_{y}(\bm{s}_{T,h}^{test})$ given posterior samples $\sigma$ and $\psi_{x}(\bm{s})$ and $\psi_{y}(\bm{s})$.
For Neza, the estimated bivariate CRPS's for the 80 test pairs are 2.947 ($\Sigma$ constant), 2.624 ($\phi^{*}=30$), 2.636 ($\phi^{*}=10$) and 2.962 ( $\phi^{*}=1$). The spatially varying kernel models with $\phi^{*}=30$ and $\phi^{*}=10$ show similar performance, being preferred to the spatially varying kernel model with $\phi^{*}=1$ and constant kernel model.
For Belo Horizonte, $H=2,625$ (50$\%$ of the total number of points) are randomly held out for the testing dataset.
The estimated bivariate CRPS for the 2,625 test pairs are 156.43 ($\Sigma$ constant), 152.59 ($\phi^{*}=30$), 152.29 ($\phi^{*}=10$) and 152.64 ($\phi^{*}=1$).  The spatially varying kernel models are indistinguishable and slightly outperform the constant kernel model.

\begin{figure}[htbp]
  \begin{center}
   \includegraphics[width=15cm]{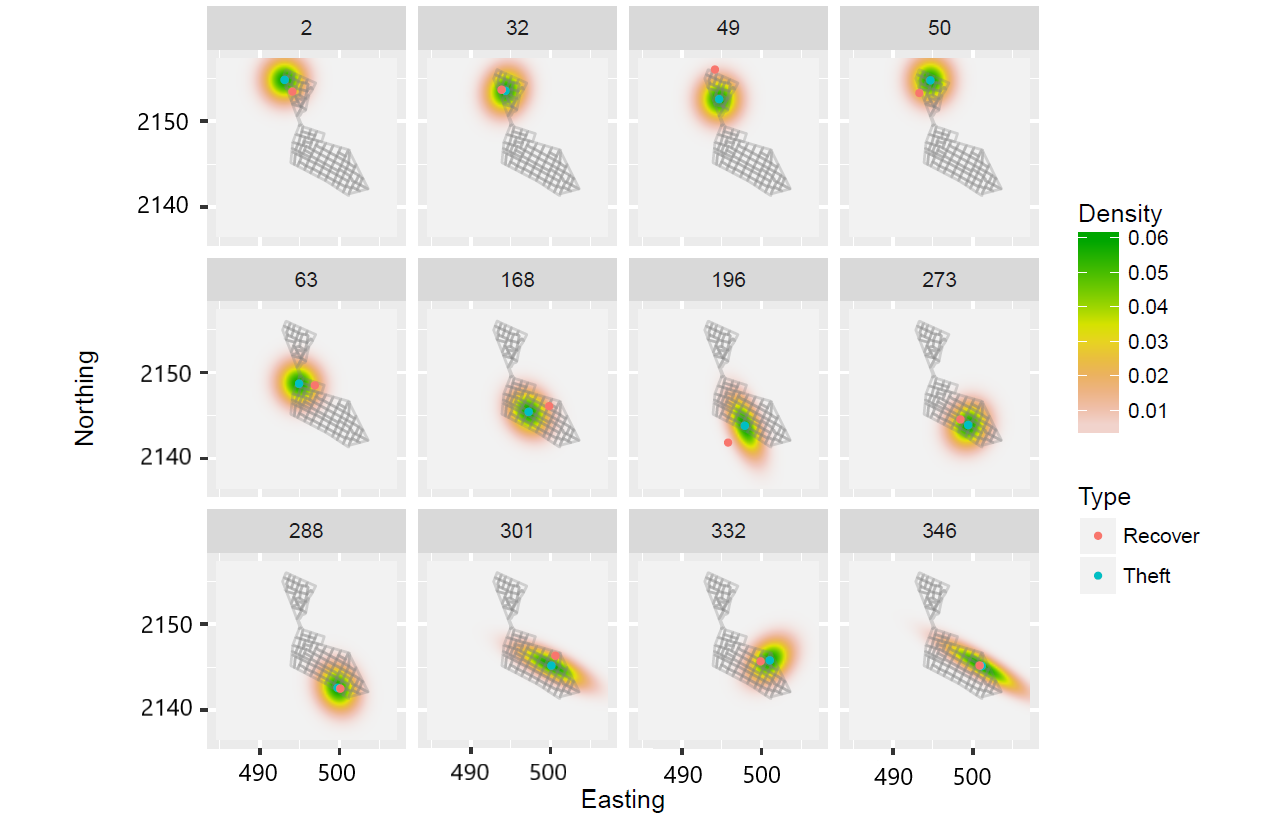}
  \end{center}
  \caption{Predictive conditional density $f_{R}(\cdot|\cdot)$ for selected pairs in Neza with $\phi^{*}=30$.}
  \label{fig:CDNZ}
\end{figure}

\begin{figure}[htbp]
  \begin{center}
   \includegraphics[width=15cm]{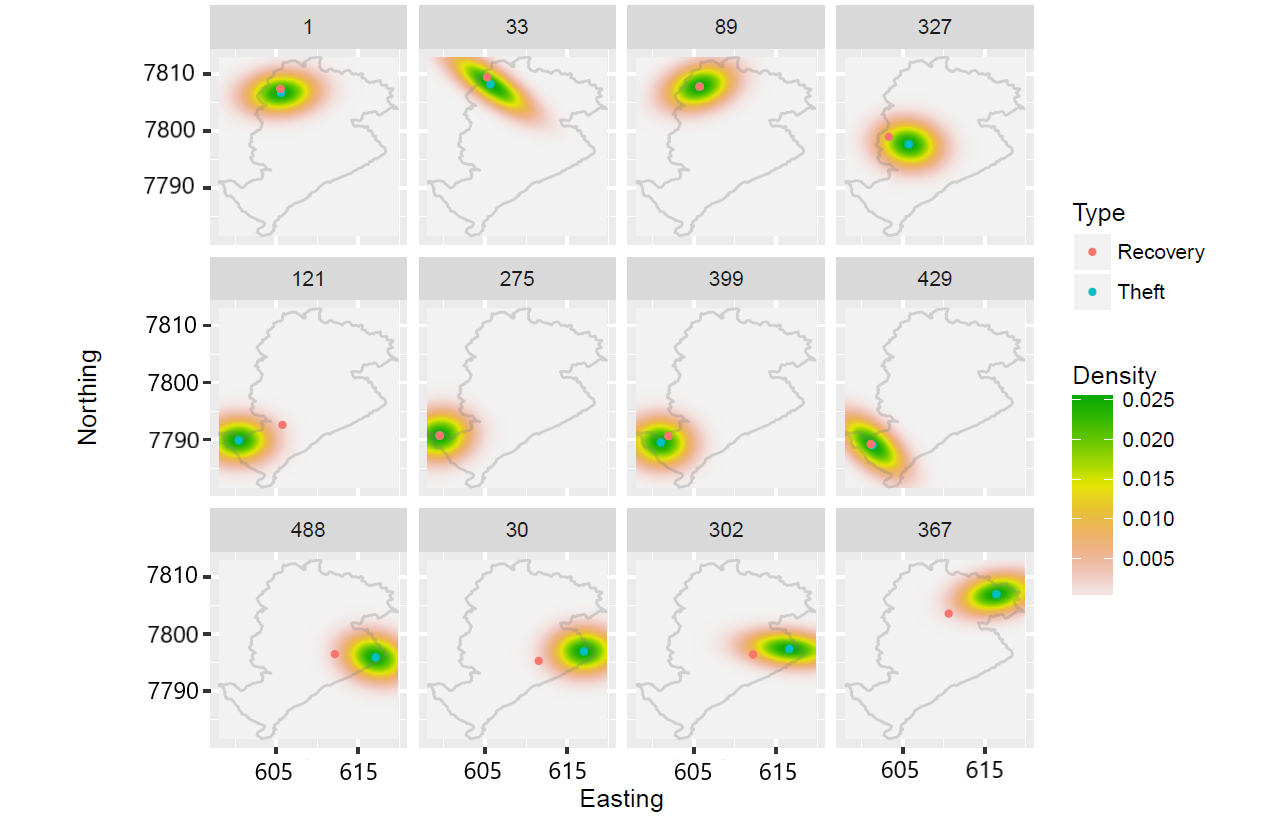}
  \end{center}
  \caption{Predictive conditional density $f_{R}(\cdot|\cdot)$ for selected pairs in Belo Horizonte with $\phi^{*}=10$.}
  \label{fig:CDBH}
\end{figure}

Some illustrative posterior predictive densities for recovery location given theft location are presented for the testing dataset of recovery locations.  Specifically,
Figure \ref{fig:CDNZ} shows the conditional density $f_{R}(\cdot|\cdot)$ defined in (\ref{eq:CD}) for some pairs in Neza
(using $\phi^{*}=30$, which, above, gave the best predictive performance).
For ID 2, 32, 49 and 50, theft locations are in the north region, and their conditional densities are close to be uncorrelated densities.
On the other hand, conditional densities for some pairs in the south regions, e.g., ID 196, 301, 332 and 346, show different shapes for the kernels.
Figure \ref{fig:CDBH} shows the conditional density $f_{R}(\cdot|\cdot)$ for some pairs in Belo Horizonte (under $\phi^{*}=10$, which gave the best predictive performance).
Conditional densities across pairs, e.g., ID 33, 302 and 429, show different shapes for the kernels.
These results suggest that the shapes of conditional densities are location dependent, particularly showing sensitivity when the theft locations are near the boundary of the region.

\section{Joint Point Pattern Modeling}
\label{sec:Joint}
Here, linking the theft location point pattern and the recovery location point pattern is considered.  As noted in the Introduction, we find ourselves in an origin-destination setting but at point-referenced scale rather than areal unit scale.  We build a \emph{joint}  intensity of the form $\lambda(\bm{s}_{o}, \bm{s}_{d})$ over pairs of locations $(\bm{s}_{o}, \bm{s}_{d}) \in D_{o} \times D_{d}$ where $\bm{s}_{o}$ is a theft location and $\bm{s}_{d}$ is a recovery location.
A marginal intensity surface for both theft locations and for recovery locations emerges.
In addition, to attempt to understand the flow of vehicles from theft location to recovery location, we consider a theft neighborhood, say $B_{o} \subset D_{o}$ and a recovery neighborhood say $B_{d} \subset D_{d}$.  Then, we can ask for the incidence (or chance) of theft in $B_{o}$ with recovery in $B_{d}$.  In fact, $D_{d}$ can be partitioned into several neighborhoods to see the flow from $B_{o}$ into each.

Now, a LGCP is introduced for \emph{pairs} of locations as a joint point process model over $D_{o} \times D_{d} \subset \mathbb{R}^{2}\times \mathbb{R}^{2}$.  In fact, $D_{o}=D_{d}=D$ is taken in the sequel.
We denote observed pairs as $\mathcal{S}_{P}=\{\bm{s}_{P,1}, \ldots, \bm{s}_{P,m} \}=\{(\bm{s}_{R,1}, \bm{s}_{T,1}^{*}), \ldots, (\bm{s}_{R,m}, \bm{s}_{T,m}^{*}) \}$; $R$ denotes recovery, $T$ denotes theft.
The intensity function for observed pairs is specified as
\begin{align}
\log \lambda(\bm{s}_{R}, \bm{s}_{T}^{*})&=\bm{X}_{R}(\bm{s}_{R})\bm{\beta}_{R}+\bm{X}_{T}(\bm{s}_{T}^{*})\bm{\beta}_{T} \nonumber \\
&+\eta (\bm{s}_{R}-\bm{s}_{T}^{*})'\Sigma(\bm{s}_{T}^{*})^{-1}(\bm{s}_{R}-\bm{s}_{T}^{*})+z_{R}(\bm{s}_{R})+z_{T}(\bm{s}_{T}^{*}),  \label{pairsintensity}
 \\
\bm{z}_{R}&  \sim \mathcal{N}(\bm{0}, \mathbf{C}_{\bm{z}_{R}}), \quad \bm{z}_{T} \sim \mathcal{N}(\bm{0}, \mathbf{C}_{\bm{z}_{T}}).
\end{align}
Here, $z_{R}(\bm{s})$ and $z_{T}(\bm{s})$ are mean $0$ GP's with covariance functions $C_{R}$ and $C_{T}$, respectively. $\bm{z}_{R}=(z(\bm{s}_{R,1}), \ldots, z(\bm{s}_{R,m}))$, $\bm{z}_{T}=(z(\bm{s}_{T,1}^{*}), \ldots, z(\bm{s}_{T,m}^{*}))$, and $\mathbf{C}_{\bm{z}_{R}}=[C_{R}(\bm{s}_{R,i},\bm{s}_{R,j})]_{i,j=1, \ldots, m}$ and $\mathbf{C}_{\bm{z}_{T}}=[C_{T}(\bm{s}_{T,i}^{*},\bm{s}_{T,j}^{*})]_{i,j=1, \ldots, m}$.
Exponential covariance functions are assumed for $C_{R}$ and $C_{T}$, i.e., $C_{R}(\bm{u}, \bm{u}^{'})=\sigma_{R}^2\exp(-\phi_{R}\|\bm{u}-\bm{u}^{'}\|)$ and $C_{T}(\bm{u}, \bm{u}^{'})=\sigma_{T}^2\exp(-\phi_{T}\|\bm{u}-\bm{u}^{'}\|)$.

The first two terms of the log intensity introduce recovery location and theft location covariates. The third term introduces a distance between theft location and recovery location, analogous to ``commuting distance'' in customary origin-destination modeling.  It takes a local (spatially varying) form through $\Sigma(\bm{s}_{T}^{*})$, the spatially varying kernel presented in the previous section. $\eta$ is the \emph{critical} parameter; it captures the dependence between the point patterns.  If it is not significant, then the joint intensity factors into an intensity for the theft point pattern times an intensity for the recovery point pattern.  In fact, $\eta$ is expected to be negative, i.e., the recovery locations tend to be observed nearer the corresponding theft locations. In addition, the local $\Sigma(\bm{s}_{T}^{*})$ enables directional preference for $\bm{s}_{R}$ near the boundaries of $D$.  The fourth and fifth terms provide recovery location and theft location random effects using Gaussian processes.  Without them, the analogue of an NHPP is available; with them, we have the analogue of a LGCP.   This joint specification only employs the complete pairs in the data and will need a large number of pairs in order to learn about the local random effects adjustments.

As above, the likelihood is approximated by gridding $D$ into $K$ blocks.   Now, employing $K \times K$ grid for $D \times D$, we obtain
\begin{align}
\mathcal{L}(\mathcal{S}_{P})&\propto \exp\biggl(-\int_{D}\int_{D}\lambda(\bm{u}_{R}, \bm{u}_{T})d\bm{u}_{T}d\bm{u}_{R} \biggl)\prod_{j=1}^{m}\lambda(\bm{s}_{R, j}, \bm{s}_{T,j}^{*}) \\
&\approx \exp\biggl(-\sum_{k=1}^{K}\sum_{k^{'}=1}^{K} \lambda(\bm{u}_{R,k}, \bm{u}_{T,k^{'}})\Delta_{T, k}\Delta_{T, k^{'}} \biggl)\prod_{k=1}^{K}\prod_{k^{'}=1}^{K}\lambda(\bm{u}_{R, k}, \bm{u}_{T, k^{'}})^{n_{kk^{'}}}
\end{align}
where $\sum_{k=1}^{K}\sum_{k^{'}=1}^{K}n_{kk^{'}}=m$.

\subsection{Results}
\label{sec:ResultJoint}

We demonstrate results only for the Belo Horizonte data because of the large number of pairs of points (again, 5,250 points).  The small number of pairs for the Neza region (only $68$ in the South) precludes informative model fitting for \eqref{pairsintensity}.
Without covariates, the intensity model for the Belo Horizonte data becomes
\begin{align}
\log \lambda(\bm{s}_{R}, \bm{s}_{T}^{*})&=\beta_{0}+\eta (\bm{s}_{R}-\bm{s}_{T}^{*})'\Sigma(\bm{s}_{T}^{*})^{-1}(\bm{s}_{R}-\bm{s}_{T}^{*})+z_{R}(\bm{s}_{R})+z_{T}(\bm{s}_{T}^{*}), \\
\bm{z}_{R}(\mathcal{S}_{R})&  \sim \mathcal{N}(\bm{0}, \mathbf{C}_{R}(\mathcal{S}_{R},\mathcal{S}_{R})), \quad \bm{z}_{T}(\mathcal{S}_{T}^{*}) \sim \mathcal{N}(\bm{0}, \mathbf{C}_{T}(\mathcal{S}_{T}^{*},\mathcal{S}_{T}^{*})).
\end{align}
$K=305$ grids are taken, the as same as in conditional density specification.
Two models are fitted: (i) a LGCP without the spatially varying distance measure (LGCP-Ind, i.e., $\eta=0$) and (ii) a LGCP with this measure (LGCP-Dep).  For priors we assume $\sigma_{R}^2, \sigma_{T}^2\sim \mathcal{IG}(2, 0.1)$, $\beta_{0},\eta \sim \mathcal{N}(0, 100)$ and $\phi_{R}, \phi_{T}\sim \mathcal{U}[0, 10]$.
For sampling $\beta_{0}$ and $\eta$, an adaptive random walk MH algorithm is implemented.
Elliptical slice sampling is adopted for $\psi_{x}(\bm{s})$, $\psi_{y}(\bm{s})$, $z_{R}(\bm{s})$ and $z_{T}(\bm{s})$. We fixed the tuning parameter $\phi^{*}=1$.
For high dimensional grids, approximate Gaussian process models can provide efficient process sampling, e.g., nearest neighbor Gaussian processes \citep{Dattaetal(16a)} and multi-resolution Gaussian processes \citep{Katzfuss(17)}.
\cite{ShirotaBanerjee(18)} propose scalable inference for Gaussian Cox process models.
Integrated nested Laplace approximation (INLA) for the LGCP \citep{Simpsonetal(16b)} is another option.
However, the elliptical slice sampling approach is user friendly with no tuning required, enabling easy implementation for moderate point pattern sizes.
20,000 samples are discarded as the burn-in period and the subsequent 20,000 samples are retained as posterior samples.
The likelihood value for the LGCP-Dep (-10789 [-10829, -10744]) is much larger than that of the LGCP-ind (-14348 [-14381, -14319]), demonstrating the superiority of LGCP-Dep.
Furthermore, the estimated value of $\eta$ is significantly negative (-0.044 [-0.046, -0.043]), as expected.

\begin{figure}[htbp]
  \begin{center}
   \includegraphics[width=15cm]{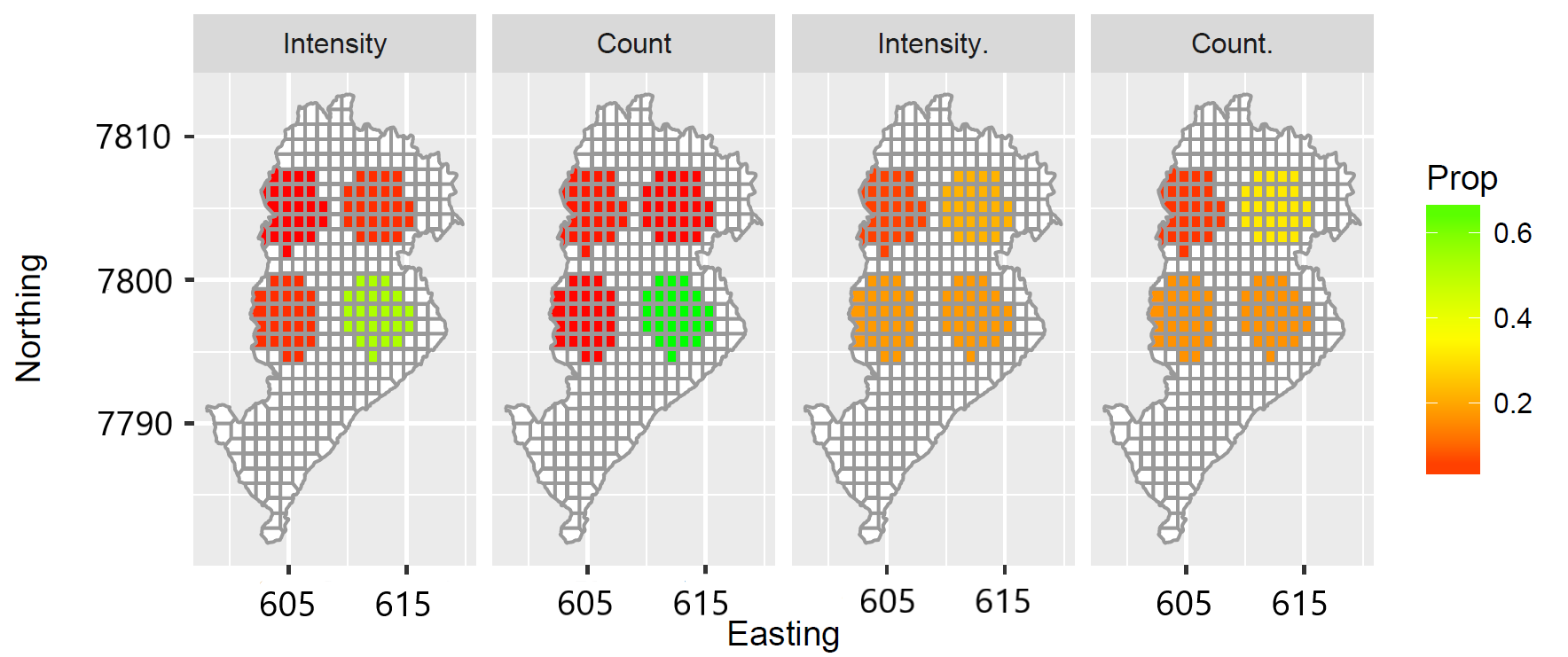}
  \end{center}
  \caption{Proportion of testing data and predictive intensities using four subregions for the theft regions around $L_{1}$ (left two panels: southeast region) and $L_{2}$ (right two panels: northeast region).}
  \label{fig:JointPreBH}
\end{figure}

We demonstrate the predictive flow from theft locations to recovery locations.
Four subregions are created, each of  which is composed of $G=25$ grid cells, around four locations: $L_{1}=(612.5, 7797.5)$, $L_{2}=(612.5, 7805)$, $L_{3}=(605, 7797.5)$ and $L_{4}=(605, 7805)$.
The proportion of predictive intensities and counts are compared for the same theft subregion, i.e., $p_{int}(B_{d}|B_{o})=\frac{\lambda(B_{d}, B_{o})}{\lambda(D_{d}, B_{o})}$  and $p_{count}(B_{d}|B_{o})=\frac{N(B_{d}, B_{o})}{N(D_{d}, B_{o})}$ where $B_{d}\subset D_{d}$ and $B_{o}\subset D_{o}$ and the $\lambda$s are integrated intensities.
Figure \ref{fig:JointPreBH} looks at two origin regions. The left two panels are associated with the southeast origin region, a high intensity region.  The results show that the flow is highly concentrated in that region, in agreement with the actual recoveries in the testing dataset. The right two panels are associated with the northeast region, a lower intensity region.  The results there show more flow from that region to the other three regions, again in agreement with the recoveries in the testing dataset.
Hence, we see that the nature of the concentration of recovery locations is dependent on theft location and that our model is able to capture this dependence.

\section{Summary and Future Work}
\label{sec:Summary}
We have considered a little-studied problem for point patterns namely the setting where we have a point pattern of origins over $D \subset \mathbb{R}^2$ (in our case, locations of car thefts) and an associated partial point pattern of destinations, again over $D \subset \mathbb{R}^2$ (in our case locations of car recoveries). We have considered three issues.  The first seeks to learn about the point pattern of theft locations using an NHPP and a LGCP model.  The second question looks at prediction of the recovery location given the theft location, adopting  a geostatistical conditional regression specification.  The third question investigates the dependence between the theft location point pattern and the recovery location point pattern.  The point patterns are viewed as providing origin-destination pairs of points and a joint intensity model is supplied over a subset of $\mathbb{R}^2 \times \mathbb{R}^2$.

A potential follow-on analysis here would return to Section 4 and the partially observed recoveries.
When an associated recovery location is available under this model, we can mark the location with a `1'.  On the other hand, all of the theft locations without a recovery location receive a mark of `0'.  This opens up the possibility of preferential sampling \citep[][]{Diggleetal(10)}.  The question of whether the theft location influences the probability of recovery can be examined.

It is worth emphasizing that our approaches here can be applied to other origin-destination problems where the origins and destinations are provided at point level, i.e., as geo-coded locations.  We have noted that working at the highest spatial resolution provides a more clear picture than working at areal unit scales with regard to the origin surface, the destination surface, and the dependence between the surfaces.  Furthermore, with larger datasets, the dependence might be also included between the $z_{R}(\bm{s}_{R})$ process and the $z_{T}(\bm{s}_{T})$ process through say coregionalization \citep[][]{BanerjeeCarlinGelfand(14)}. This would add illumination to the dependence structure between the two surfaces.

One path for future work will investigate a much different application.  We will examine economic labor force data where, for an individual, we have the location where she/he resides as well as the location where she/he works.  Working with metropolitan areas will provide much larger point patterns with much more demanding model fitting.  Future work with theft-recovery data would introduce consideration of time, i.e., we will have not only the location of the theft but also the time of the theft.  Similarly, we have not only a location for the recovery but as well the time of the recovery, with an implicit order in time and a time lapse for the latter relative to the former.  Unfortunately, at present, neither of the datasets provide time information needed to enable such investigation.

\section*{Acknowledgements}
\label{sec:Ack}
The authors thank Renato Assun{\c{c}}{\~{a}}o for providing the datasets and shape file of Belo
Horizonte and the Nezahualcoyotl Town Hall, Arturo Arango, and Direccion General de Seguridad Ciudadana (in Mexico) for providing the Neza dataset. The authors also thank the reviewers for valuable comments, in particular for suggesting links to the literature, which consequentially improved the manuscript.
This work of the second author was partially funded by Grant MTM2016-78917-R from the Spanish Ministry
of Science and Education, and Grant P1-1B2015-40 from University Jaume I. The
work of the first author was supported in part by the Nakajima Foundation. The computational results are obtained by using Ox version 7.1 \citep{Doornik(07)}.

\small{
\bibliographystyle{chicago}
\bibliography{OAD}
}

\end{document}